\documentclass[conference]{IEEEtran}
\IEEEoverridecommandlockouts
\usepackage{cite}
\usepackage{amsmath,amssymb,amsfonts}
\usepackage{algorithmic}
\usepackage{graphicx}
\usepackage{textcomp}
\usepackage{url}
\usepackage{standalone} % prepare tikz pictures as standalone
\usepackage{tikz}
\usepackage[hidelinks]{hyperref}
\usepackage{xcolor}
\usepackage{relsize}
\usepackage{subcaption}

\definecolor{myyellow}{HTML}{ffeb3b}
\definecolor{mygreen}{HTML}{8bc34a}
\definecolor{myblue}{HTML}{4f61e5}
\definecolor{myred}{HTML}{f44336}
\definecolor{myviolet}{HTML}{9c27b0}
\definecolor{myteal}{HTML}{009688}
\newcommand{\figref}[1]{\figurename~\ref{#1}}

\newcommand{\secref}[1]{Sect.~\ref{#1}}

\usepackage{fontawesome5} %% for Orcid logo
\newcommand{\orcid}[1]{\href{https://orcid.org/#1}{\textcolor[HTML]{A6CE39}{$^{\textrm{\faOrcid}}$}}}

\usepackage{listings}
\lstdefinestyle{customc}{%
  backgroundcolor = \color{lightgray!30!white},
  frame=ltbr,framesep=4pt,framerule=0pt,
  xleftmargin=4pt,
  xrightmargin=4pt,
  belowcaptionskip=1\baselineskip,
  breaklines=true,
  language=bash,
  showstringspaces=false,
  basicstyle=\scriptsize\ttfamily,
  keywordstyle=\bfseries\color{blue},
  keywordstyle=[2]\bfseries\color{green!60!black}, % data types
  keywordstyle=[3]\fontfamily{lmtt}\fontseries{b}\selectfont\color{myblue},  % wd
  keywordstyle=[4]\fontfamily{lmtt}\fontseries{b}\selectfont\color{mygreen}, % cgroupid
  keywordstyle=[5]\fontfamily{lmtt}\fontseries{b}\selectfont\color{myred},   % taskid
  keywordstyle=[6]\fontfamily{lmtt}\fontseries{b}\selectfont\color{myviolet}, % pid
  keywordstyle=[7]\fontfamily{lmtt}\fontseries{b}\selectfont\color{myteal}, % pid
  numberstyle=\tiny,
  commentstyle=\itshape\color{purple!60!black},
  identifierstyle=\mdseries\color{black},
  keywords={for,if,while}, % list explicitly, otherwise we cannot overwrite the basic types
  morekeywords=[2]{size_t,void,int,uintw_t,uint64_t,uint32_t,__m256i,__m128i,simd8,uint8_t,UINT64_C},
  alsoletter=0123456789-_.,
  morekeywords=[3]{home,witzke,nf-rnaseq,outdir,work,52,f11191010952840e07774a95bcd36e,},
  morekeywords=[4]{131863,cgroupid},
  morekeywords=[5]{NFCORE_RNASEQ,RNASEQ,FASTQ_FASTQC_UMITOOLS_TRIMGALORE,TRIMGALORE,WT_REP2,prepare_level2,NFCORE_RNASEQ_RNASEQ_FASTQ_FASTQC_UMITOOLS_TRIMGALORE_TRIMGALORE_WT_REP2,NFCORE_RNASEQ_RNASEQ_FASTQ_FASTQC_UMITOOLS_TRIMGALORE_TRIMGALORE_},
  morekeywords=[6]{1169224,pid},
  morekeywords=[7]{pod,nf-002fdc87df831ed4f74f0f2a66482475},
  literate={cgroupid}{{\unskip\bgroup\fontfamily{lmtt}\fontseries{b}\selectfont\color{mygreen}{cgroupid}{\unskip\egroup}}}8
               {path}{{\unskip\bgroup\fontfamily{lmtt}\fontseries{b}\selectfont\color{myblue}{path}{\unskip\egroup}}}4
               {read}{{\unskip\bgroup{\fontfamily{lmtt}\fontseries{b}\selectfont{read}}{\unskip\egroup}}}4
               {open}{{\unskip\bgroup{\fontfamily{lmtt}\fontseries{b}\selectfont{open}}{\unskip\egroup}}}4
               {'}{{\textquotesingle}}1
               {~}{{\centeredtilde}}1,
  escapeinside={(*@}{@*)}
}
\newcommand{\mypid}[1]{\unskip\bgroup\fontfamily{lmtt}\fontseries{b}\selectfont{\color{myviolet}{#1}}{\unskip\egroup}}
\newcommand{\code}[1]{\lstinline[basicstyle=\small\ttfamily]!#1!}

\begin{document}

\title{Low-level I/O Monitoring for Scientific Workflows
\thanks{This work received funding from the German Research Foundation (DFG), CRC 1404: \emph{FONDA: Foundations of Workflows for Large-Scale Scientific Data Analysis}}
}

\author{\IEEEauthorblockN{
    Joel Witzke\IEEEauthorrefmark{1}\orcid{0000-0002-0831-8078},
    Ansgar Lößer\IEEEauthorrefmark{2}\orcid{0000-0002-7627-9664},
    Vasilis Bountris\IEEEauthorrefmark{3}\orcid{0000-0002-8682-7302},
    Florian Schintke\IEEEauthorrefmark{1}\orcid{0000-0003-4548-788X},
    Björn Scheuermann\IEEEauthorrefmark{2}
  }
\IEEEauthorblockA{\IEEEauthorrefmark{1}\textit{Zuse Institute Berlin, Germany}}
\IEEEauthorblockA{\IEEEauthorrefmark{2}\textit{TU Darmstadt, Germany}}
\IEEEauthorblockA{\IEEEauthorrefmark{3}\textit{HU Berlin, Germany}}
\IEEEauthorblockA{witzke@zib.de, ansgar.loesser@kom.tu-darmstadt.de,
  vasilis.bountris@informatik.hu-berlin.de}
\IEEEauthorblockA{schintke@zib.de, bjoern.scheuermann@tu-darmstadt.de}
}

\maketitle

\begin{abstract}
%% 4 sentences:
%% State the problem
While detailed resource usage monitoring is possible on the low-level using proper tools, associating such usage with higher-level abstractions in the application layer that actually cause the resource usage in the first place presents a number of challenges.
%% Say why it is an interesting problem
Suppose a large-scale scientific data analysis workflow is run using a distributed execution environment such as a compute cluster or cloud environment and we want to analyze the I/O behaviour of it to find and alleviate potential bottlenecks. Different tasks of the workflow can be assigned to arbitrary compute nodes and may even share the same compute nodes.
Thus, locally observed resource usage is not directly associated with the individual workflow tasks.
%% Say what your solution achieves
By acquiring resource usage profiles of the involved nodes, we seek to correlate the trace data to the workflow and its individual tasks. To accomplish that, we select the proper set of metadata associated with low-level traces that let us associate them with higher-level task information obtained from log files of the workflow execution as well as the job management using a task orchestrator such as Kubernetes with its container management.
%% Say what follows from your solution
Ensuring a proper information chain allows the classification of observed I/O on a logical task level and may reveal the most costly or inefficient tasks of a scientific workflow that are most promising for optimization.
\end{abstract}


%\tableofcontents

\section{Introduction}

% why scientific workflows exist:
% scientific workflows are used more and more for processing scientific data in many fields
% many processing large amounts of data (cite some work)
Large-scale scientific data analysis is common in different scientific fields such as particle physics~\cite{lhc_application}, radio astronomy~\cite{radiotelescope_application}, earth observation~\cite{remotesensing_application} or bioinformatics \cite{bioinformatics_application}.
Often, the scientific data analysis and its individual processing steps, which are executed and coordinated by a scientific workflow management system (SWMS), perform significant input and output, so that their overall performance largely depends on efficient I/O.
% understanding how tasks behave in terms of resource usage can enable optimizations
But, given the ever-growing complexity of modern scientific applications, we cannot expect a one-size-fits-all storage tier hierarchy and architecture for all applications~\cite{hpcio_dagstuhl}.
Thus, gaining more insight into the task's behavior and the I/O workload it induces can reveal possible optimizations that not only decrease the overall runtime of a workflow, but also make better use of shared resources such as storage media.
% e.g. knowing the I/O pattern -> choosing a "good" storage location (HDD, SSD, tempfs, nfs / ceph)
For example, having knowledge about the I/O patterns used by a task to read or write files can help in choosing adequate storage (HDD, SSD, tempfs, NFS / Ceph, etc.) for these files~\cite{iosig}.
Detailed knowledge about a task's behavior over time can also be the foundation for accurate task and workflow models, which can be leveraged to optimize resource allocations generally~\cite{bottlemod, proactive_resource_management}.

% there are a lot of possible monitoring tools (cite some)
% all / most very low level -> difficult to connect to the abstract concept of "task" in a workflow
Although there exist numerous generic tools and frameworks for single process or system monitoring, they do not cover more abstract concepts of tasks or workflows running on the system and are thus not able to associate the low-level trace data to such upper-level entities, especially in a shared or even distributed execution environment.
% that makes it difficult to actually track the behavior of a task, especially more so in a shared system
These circumstances make it difficult to track the behavior of a workflow task and its isolated contribution to the overall system and resource usage.

% our contribution
After an introductory background section on scientific workflows (\secref{sec:background}), we make the following contributions:
\begin{itemize}
% provide approaches for low level I/O monitoring
  \item provide three approaches for detailed I/O monitoring and discuss them briefly in \secref{sec:monitoring},
% show our implementation using eBPF
  \item describe our implementation of low-level I/O monitoring using eBPF in \secref{sec:implementation},
% we show the limited ways of connecting that information to the concept of tasks (first physical, then logical); especially for nextflow where that is possible quite elegantly (although probably not thought of that way by nextflow)
  \item discuss general approaches to associate the monitoring data to tasks of scientific workflows (first to physical, then to logical tasks), highlight the emerging difficulties, and provide a solution for the Nextflow and Airflow scientific workflow management system in \secref{sec:associating-tasks}, and
% using monitoring and the means to connect that to tasks, we show I/O access patterns and statistics of those tasks (while they remain treated as black boxes)
  \item apply our monitoring solution and the method to associate the data to tasks in practice to demonstrate observed I/O behavior in tasks of well-known and widely used workflows from the nf-core workflows~\cite{nfcore} of Nextflow in \secref{sec:evaluation}.
\end{itemize}

\section{Background: Scientific Workflow Execution}
\label{sec:background}

% short inro by examples:
Many scientific fields have the need to process an ever-increasing amount of data. Those include for example the Large Hadron Collider (LHC) already retaining data in the petabyte range~\cite{lhc_application}, radio telescopes generating 160 terabytes per second~\cite{radiotelescope_application} or analyzing earth observation data~\cite{remotesensing_application}. The need to manage and automate the execution of these data processing pipelines is often addressed by scientific workflows, executed through scientific workflow management systems (SWMS).

% what is a scientific workflow:
% consists of multiple tasks chained together -> each succeeding task gets input data from its predecessors
A scientific workflow consists of logical tasks, which are chained together so that each task uses a base input and/or the results of a preceding task. Therefore, a scientific workflow can be represented as a directed acyclic graph (DAG).
% tasks are often made from a "standalone" program that covers one step of data processing, visualization, etc.
A logical task usually corresponds to one program that covers one step in the workflow, such as  (pre-)processing a certain type of data, performing a specific visualization or representation of (intermediate) results, etc.
% each logical task is executed with 1 or more physical tasks (depending on the task, input data may be split accordingly to data size, the execution environment, etc.)
Depending on the task, a logical task may be performed by multiple physical task instances during a particular execution of the scientific workflow. That may happen, for example, if a big chunk of data can be partitioned into smaller chunks, so that the same task can be executed on each partition independently and simultaneously.

% how they are executed:
Scientific workflows are typically executed using a scientific workflow management system (SWMS) such as Nextflow~\cite{nextflow}, Apache Airflow~\cite{using_airflow}, Snakemake~\cite{snakemake} or Pegasus~\cite{pegasus}, to name a few. The SWMS receives the workflow description, initial input data and configuration for the execution environment. It then schedules physical tasks when they are ready to run, meaning all preceding tasks they depend on have finished.
% execution can happen locally, using containers, or distributed (slurm, kubernetes), depends on the cluster and what the engine supports
The SWMS usually supports multiple ways to execute tasks such as simple local execution, containerized execution (using solutions such as Docker or containerd) or, highly relevant in practice, distributed executions, e.g., using Kubernetes~\cite{kubernetes} or Slurm~\cite{slurm}. Which executor is used usually depends on the SWMS configuration and the available compute cluster for workflow execution.
% usually a network (NFS) or distributed fs (ceph) is used to exchange intermediate results
To exchange intermediate results in a distributed setting, a network or distributed filesystem (e.g., NFS~\cite{nfs} or Ceph~\cite{ceph}) is often used.

% tasks are often blackboxes -> their behavior is not known; neither to the "user" of a task (that is the person building a workflow out of singular tasks), the "user" of a workflow (the person who wants to process his data with the workflow), nor to the cluster system / administrator (who is responsible for the "low level" execution of everything)
The tasks of a workflow are considered to be black boxes. The required inputs and the output files they produce are specified as part of a workflow definition, which is the actual input for the SWMS to execute a whole workflow. However, the task's particular behavior is not specified or documented in any way. It is neither necessarily known by the user of a workflow (the scientists using the workflow to process their data), nor by the administrator of the system the workflow is executed on, nor even by the workflow designers when they, for example, include a third-party program as a task in a workflow.
% sometimes resource limitations are given for tasks -> fixed maximums, often overestimated (cite), don't provide insights to their "behavior" (e.g. I/O patterns or CPU usage over tasks runtime)
As part of the workflow definition, resource limitations may be provided per task. Those limitations are fixed and enforced (e.g., using Linux control groups, short cgroups~\cite{cgroups}) over the whole runtime of the task. Since a violation of a set resource limit usually leads to the task being terminated, the limits are often set very generously, so that the actual resource requirement is regularly overestimated~\cite{task_inequality, feedback_resource_alloc, predicting_dynamic_memory}. In addition, the users often do not see a direct benefit of good resource estimates, as they are typically only accounted for the time the job actually ran. So, often the maximum job time of an execution queue is chosen for job specifications. This benefits the users because they do not need to manage estimates for each workflow submission and execution separately.
Also, resource limitations do not provide detailed insight into a task's behavior as they, for example, offer no information about CPU or memory usage over time or I/O patterns on the input and output files.

% SWMSs don't provide detailed monitoring themselves
% Nextflow Trace Report
% practically no overhead
% only aggregated information over the whole execution
Metrics gathered directly by the scientific workflow management
systems are typically coarse-grained on a task level, if gathered at
all. % directly integrated into nextflow -> no additional tools needed
% also directly connects the metrics to the tasks (instead of to low level processes which must be connected to the tasks themselves)
For example, Nextflow provides summary data for each task's
lifetime in its trace reports~\cite{towardsmonitoring}.
To get more detailed insights into the particular behavior of the tasks, additional means become necessary as we will see in the following. Especially, associating low-level measurements with the tasks of a workflow can be challenging.

\section{Detailed Low-Level I/O Monitoring}
\label{sec:monitoring}

% introduction to this section: briefly present three possible approaches
We present three methods for low-level I/O monitoring that can build the basis of our approach to associate low-level monitoring data with upper-level entities. All approaches can be implemented to produce their low-level monitoring data in a common format (see \figref{fig:log-example} for an example), which makes them exchangeable depending on the use case.

\begin{figure*}[ht]
  \includegraphics{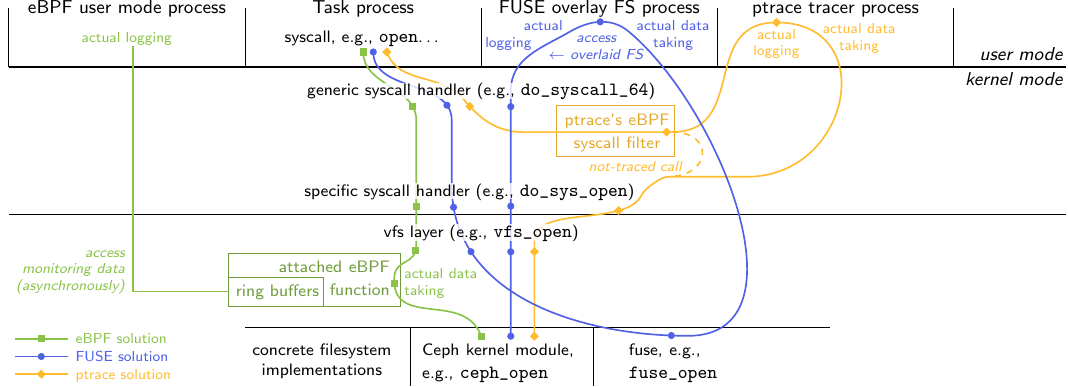}
  \caption{Components involved in different monitoring approaches.
\label{fig:monitoring}}
\end{figure*}

In order to realize detailed I/O monitoring, we are interested in the arguments of the individual read and write requests, together with some metadata. In Linux, this can be done with several different technologies, each offering different advantages and drawbacks. \figref{fig:monitoring} provides an overview of the data and control flow for the three approaches considered: Extended Berkeley Packet Filter, a FUSE overlay file system, and a ptrace-based approach.

\subsection{I/O tracing with a FUSE overlay file system}

% Intro FUSE module
Using the \emph{filesystem in user space (FUSE)} kernel module~\cite{fusedoc}, users can implement filesystems in userspace via a FUSE client. This enables the integration of various storage systems into the filesystem tree, without modifying kernel code. Popular examples include NTFS-3G~\cite{ntfs3g} and SSHFS~\cite{sshfs}.

% Data path
With FUSE, all I/O operations are issued as usual and reach the \emph{virtual filesystem (VFS)} layer~\cite{vfsdoc}. If a FUSE mount point is addressed and a change data change or data that is not in the page cache is requested, the FUSE client process is triggered via the FUSE protocol, which means communicating with the client over a special file handle. The client then handles the request, potentially providing the requested data.

% Impl of monitoring
Using this technology, an overlay filesystem can be implemented. It redirects changes to the underlying storage, while gathering detailed monitoring data akin to \emph{iofs}~\cite{iofs}, which monitors I/O less detailed.
Monitoring with FUSE is blocking, since every monitored operation needs to pass the FUSE client.
So, this approach does not bear the risk of missing a single event, but may slow down the application itself.
One can configure to log with or without the influence of the page cache. Logging \emph{with} page cache effects resembles monitoring of the block device while logging \emph{without} page cache effects results in monitoring of the exact I/O calls of the monitored application but may increase the overhead in latency and CPU time~\cite{fuseperf}.
% Blocking and performance
The latency of an operation depends on the speed of the underlying storage, the monitoring system, and the overhead introduced by FUSE. Therefore, it is not an ideal solution for implementing a passive monitoring system that tries to minimize runtime overhead and side-channel influence. With the recent Linux kernel version 6.9 these negative effects can be mitigated for files that do not need to be monitored using data passthrough.%~\cite{someref}. %, but this is not possible for the target data.

% Injection into workflows
On distributed, containerized infrastructures, it is typical to mount a filesystem from inside the container that performs the task execution. This imposes the challenge that outside the container, the target filesystem may not be mounted and monitoring is not possible. Thus, the FUSE client must run inside the container along the monitored task. Using FUSE inside containers is generally possible by passing through the FUSE file descriptor to enable the client to communicate with the kernel-side. To adhere to the black-box approach and not change the workflow task or its container, the FUSE client should not rely on any other files such as shared libraries as many workflow containers only including the barely needed packages and files. A statically linked executable, which resides in a shared filesystem can be used to mount the overlay filesystem in the containers. Still, this approach needs adaptation in the executing infrastructure and the workflow manager of the task.

\subsection{I/O tracing with ptrace}

% Intro ptrace module
Ptrace is a system to debug processes in POSIX systems by tracing and potentially manipulating a process~\cite{ptraceman}. The managing user-level process is the tracer, inspecting one or multiple target processes, called tracees. Ptrace works with a special set of system calls and enables the tracer to set traps for different types of events, including single-step execution, entering or leaving a system call and signals arriving at the tracee. When this happens, the tracee is stopped until the tracer has received the event and resumed the tracee. The tracer can also manipulate the tracee, e.g., by changing the contents of the memory or suppressing or redirecting a signal. Ptrace is mainly used by debugging tools such as \emph{gdb} and \emph{strace}.

% Impl of monitoring
In order to implement detailed I/O monitoring, ptrace can be used to trace the root process of a task and all of its child processes. Monitoring the children can be automated by tracing the \code{fork} and \code{clone} system calls. By setting a trap for the event of entering a system call, all I/O operations can be monitored by reading the registers and memory of the tracee at the moment the system call happens. If the result of the operation is also needed, the same strategy can be applied on system call exits. When implementing this strategy, it is important to watch every relevant system call. This can be tricky, especially since new relevant system calls are introduced quite frequently. One example of this is the new asynchronous \emph{io\_uring} API~\cite{didona2022understanding}, which is too complex to parse correctly in this scenario. Instead, the allowed subset of system calls on the system can be restricted using the \emph{seccomp} module~\cite{seccompdocs}, which was designed to encapsulate the process in a secure computing environment. This might alter the behavior of the system under observation making the monitoring layer not transparent anymore and may change the result of the application. Advantageous, the ptrace method can deliver the most detailed application monitoring including direct I/O operations with the state of the complete memory of the tracee if needed.

% Blocking, data path and performance
Similar to FUSE, monitoring with ptrace is blocking, resulting in the same properties discussed above.
Synchronously redirecting every single system call and every signal to the tracer can cause a high overhead. The parameters and results of system calls have to be read by the tracer by reading registers and memory in a different userspace. This increases the memory management and communication overhead additionally. Filtering the actually monitored system calls to the relevant ones with the Linux \emph{seccomp} module (see above) can substantially reduce the overhead.
%While this is typically too restrictive for a general-purpose execution of a workflow task, it also allows a Berkeley Packet Filter (BPF) to decide whether a system call should be allowed, prevented, or passed to the tracer.
By using an eBPF filter (as shown in \figref{fig:monitoring}), the decision to pass the execution to the tracer can be based on the issued system call and some of its arguments. All other system calls can be directly handled by the kernel with neither pausing the thread nor informing the tracer, resulting in near native performance.

% Injection into workflows
Running the tracer inside a container of a task imposes similar challenges as running the FUSE client inside a container. In this case, redirecting file handles is not needed and the required capabilities (\code{CAP_SYS_PTRACE}) are often set by default. The optimization using \emph{seccomp} and an eBPF filter needs \code{CAP_SYS_ADMIN}, which can be problematic but circumvented by setting the \code{PR_SET_NO_NEW_PRIVS} bit preventing privilege escalation of child processes.
% by, e.g., using the \code{exec} system call family overriding techniques such as the suid bit.
%While this technically changes the observable behavior of the system, it seems not to be used by the tasks of scientific workflows.

\subsection{I/O tracing with extended Berkeley Packet Filter (eBPF)}

% history: network
The Berkeley Packet Filter (BPF) was introduced about 30 years ago to dynamically handle network packets directly inside the Linux kernel~\cite{cbpf}. A function, written for a special BPF architecture, provided by an application was attached along the fixed packet handling pipeline of the kernel and was interpreted and executed in the kernel context.
% then architecture revamped (cBPF -> eBPF) and architecture put to use to trace arbitrary kernel functions, not only network stuff
With the \emph{extended Berkeley Packet Filter, eBPF} the concept was generalized and so-called eBPF functions could be attached to arbitrary functions inside the kernel~\cite{ebpf_origin}, which caused additional popularity of the approach. A just-in-time compiler and virtual machine are used in the Linux kernel to accomplish this functionality, which can be used for detailed low-level monitoring of any kernel function.
% eBPF description
% when code is executed
Restricting oneself to eBPF functions on system calls (short syscalls) or tracepoints only, reduces the risk of misinterpretation of function arguments in eBPF code as such interfaces remain more stable across different kernel versions.

% what can be done
Attached eBPF code can access arbitrary memory, which allows traversal and inspection of data structures such as file handles and filesystem structures within the kernel's address space. The eBPF code itself can use its own global variables and data structures such as arrays and hashmaps, allowing state awareness of eBPF functions.
Ringbuffers on shared memory are used for communication with the usermode process that attached the eBPF functions in the first place. The usermode process can transmit information by directly manipulating the eBPF code's global variables, arrays, and hashmaps.

% restrictions
As eBPF code is executed within the kernel, there are several restrictions to protect system stability.
These restrictions are checked by a verifier embedded in the kernel before attaching the eBPF code. For implementing an I/O monitoring solution, the following restrictions are most relevant:
\paragraph*{a) Termination Guarantee} eBPF code is executed just-in-time right before or after the kernel function it is attached to. As it should have access to the environment memory at that time, the thread's execution is halted for the eBPF code execution. For system stability, the eBPF code is proven to terminate and cannot hang in infinite loops or wait for resources or communication.
    Therefore, eBPF code and its controlling usermode process communicate asynchronously over shared memory.
    Additionally, loops in eBPF code need to be restricted:
    Linux kernels before version 5.3 did not allow loops at all. A restriction circumventable by using fixed-length loops
    unrolled using a preprocessor directive in the C code compiled to eBPF code. Kernel version 5.3 supports loops if the eBPF verifier inside the kernel can statically determine the loop to terminate under all circumstances.
\paragraph*{b) Restricted Code Execution} Only eBPF virtual machine code can be executed. Calling other kernel functions or performing I/O is prohibited. To save logs or act based on the observed calls, data has to be transmitted to the controlling usermode process, which then can perform the action.
\paragraph*{c) Privilege Requirement} As eBPF code can read arbitrary kernel memory, including sensitive information, root privileges or the \code{CAP_BPF} capability are required.
%\end{itemize}
\medskip

% general use cases
Despite the restrictions, eBPF enables powerful debugging~\cite{ebpf_debugging}, security tools~\cite{ebpf_runtime_security}, and monitoring of application or system metrics, e.g., with the tools provided by BCC~\cite{bcc}.

% Impl of monitoring
For I/O monitoring, eBPF allows monitoring on the same level as the ptrace solution by tracing the I/O relevant system calls. For our implementation, we chose to attach to relevant functions of the virtual file system (VFS) layer.

% Blocking, data path and performance
The most notable difference to the FUSE and ptrace approaches presented above is the lack of context switches using a pure eBPF solution. The execution flow of the application(s) under monitoring is only slightly detoured by the eBPF component inside the kernel and does not synchronously enter a different user-mode process again (see \figref{fig:monitoring}). This makes the eBPF approach the most lightweight option for the monitored application(s). Since the approach works asynchronously and does not switch contexts, timestamps at the data taking are most accurate, meaning as close as possible to the actual time the task requested the I/O operation.

% Injection into workflows
With this approach, the data taking happens inside the kernel. It is compatible with container-based execution of workflows, as containers do not virtualize their own kernel and there is just one kernel per physical node. Thus, every I/O request passes through the same kernel that has the monitoring eBPF code attached. The usermode process that takes the data from the ringbuffers and writes it to a file can be executed anywhere: inside the task's container, in its own container or directly on the host. Of the presented approaches, eBPF most seamlessly lets inject the monitoring into the workflow and tasks while treating them as black boxes.

\section{I/O Monitoring Using eBPF in Detail}
\label{sec:implementation}

The previous section discussed three different I/O monitoring approaches in general. Now, we discuss the I/O monitoring of processes with eBPF and our implementation in detail.

\subsection{The eBPF usermode process and eBPF function}

% our solution
We use Python with the \emph{BPF Compiler Collection (BCC)}~\cite{bcc} for the usermode monitoring process.
With BCC, the eBPF code can be written in C and is then just-in-time compiled to eBPF to be passed to and attached to the kernel. The Python code handles parsing and processing of the commandline arguments, prepares the eBPF code accordingly, loads it and then retrieves and prints event messages, i.e., monitoring data, transmitted from the eBPF code. We log every read or write as well as file open, close and deletion. Filtering these logs (an example is given in \figref{fig:log-example}) is possible by event type, process id, user or group of the process that does the action, and the directory of the accessed file.

% ansatz
The eBPF code is attached to functions of the virtual file system (VFS) layer inside the kernel such as \code{vfs_open}, \code{vfs_read}, \code{vfs_write} and \code{vfs_close}. Code is executed on both entry and exit of the function to allow capturing of a call's arguments as well as its return value.
%% moved from above
The VFS layer resides between the functions handling system calls and the actual filesystems. It is the deepest generic (filesystem-agnostic) layer in the Linux kernel processing I/O and well suited for monitoring, as all I/O calls pass through it. Additionally, interesting logic like resolving relative paths and symbolic links are done at that point, making monitoring and tracing less tedious.
% problem with vfs_open
The VFS functions are \emph{usually} invoked on all I/O calls. An exception on file open and create operations is \code{vfs_open}. In specific cases and depending on the filesystem, \code{atomic_open} is tried before. On success, the file is opened and \code{vfs_open} is skipped. An easy to miss fact.

% file handles
Opened files are represented in the VFS layer by a pointer to a \code{file} structure---a file handle. After a file is closed the memory of its \code{file} structure will be reused for another newly opened file. Although completely idenpendent, two file handles use the same memory for their \code{file} structure (not at the same time). To simplify the analysis of the monitoring data, our eBPF code creates a unique handle id for each opened file handle using a global counter. This is done by creating an entry in a map structure inside the eBPF code that maps from the \code{file} pointer to the unique handle id on open and deleting it on close. The unique handle id is then used in the logs to represent file handles.

% inodes
% what are they and why we need them
An inode identifies a file in a filesystem and carries meta data about it. Each inode is identified by an inode id, uniquely identifying the file. Knowing the inode is important to identify hard links, meaning a file that has more than one location in the filesystem's directory hierarchy. It also aids comfortably identifying files that were moved between two open calls.
% same idea of "uniqueification" as with file handles
An inode is removed if there is no path left in the filesystem referencing it and no handles to the file are open anymore. However, inodes are reused akin to file structures and we apply similar means to avoid any confusion due to such an inode reuse, by assigning unique inode ids, which are then used for logging purposes instead of the `raw' filesystem inode ids.

% only log reads and writes from files where the open was logged as well
Read, write and close operations are only processed by the eBPF component, and then subsequently logged if the \code{open} (or \code{atomic_open}) call corresponding to the file handle was also processed. Accesses to files opened before the monitoring was started are therefore discarded. This is implemented by checking the \code{file} pointer is member in our unique handle id map when the code for an action such as a file read is executed. If that is not the case, the code will just return without taking any actions or gathering data. This logic allows to restrict the monitoring on \code{open} to files we are interested in, keeping the impact on not-monitored files and the communication to the usermode process small.

% head and example of log entry
\begin{figure*}[t!h]
%\captionsetup{skip=0pt}
\begin{subfigure}[t]{\textwidth}
\captionsetup{skip=-3pt}
\begin{lstlisting}[style=customc]
# eBPF and FUSE-Overlay-FS IO tracing log with: time_start, time_end, (*@\mypid{\emph{pid}}@*), utime_start,utime_end,stime_start,stime_end,inode,type,result,handle,offset,size,flags,path
# example open log entry:
1714067937.744, 1714067937.744, 1169224, 27151.124, 27151.124, 27151.124, 27151.124, 5277, O, 0, 35625, 0, 0, 0x00008000, /home/witzke/nf-rnaseq/outdir/work/52/f11191010952840e07774a95bcd36e/WT_REP2_1_val_1.fq.gz
# example read log entry:
1714067937.745, 1714067937.745, 1169224, 27151.124, 27151.124, 27151.124, 27151.124, 5277, R, 512, 35625, 1034, 512, 0x00008000,
\end{lstlisting}
\caption{eBPF and FUSE-Overlay-FS I/O trace example.\label{fig:log-example}}
\end{subfigure}\medskip

\begin{subfigure}[t]{.36\textwidth}
\captionsetup{skip=-3pt}
\begin{lstlisting}[style=customc]
# eBPF PID tracing log with:
# time, parent pid, (*@\mypid{\emph{pid}}@*), cgroupid
1714067937.409, 1168419, 1169224, 131863
\end{lstlisting}
\caption{eBPF PID and cgroupid trace example.\label{fig:ebpf-cgroupid-trace}}
\end{subfigure}\hfill
\begin{subfigure}[t]{.62\textwidth}
\captionsetup{skip=-3pt}
\begin{lstlisting}[style=customc]
Apr-25 19:59:04.446 [Task monitor] DEBUG n.processor.TaskPollingMonitor - Task completed > TaskHandler[id: 6; name: NFCORE_RNASEQ:RNASEQ:FASTQ_FASTQC_UMITOOLS_TRIMGALORE:TRIMGALORE (WT_REP2); status: COMPLETED; exit: 0; error: -; workDir: /home/witzke/nf-rnaseq/outdir/work/52/f11191010952840e07774a95bcd36e]
\end{lstlisting}
\caption{Example Nextflow log file entry.\label{fig:nextflow-trace}}
\end{subfigure}\medskip

\begin{subfigure}[t]{1\textwidth}
\captionsetup{skip=-3pt}
\begin{lstlisting}[style=customc, literate=
               {pod}{{\unskip\bgroup{\fontfamily{lmtt}\fontseries{b}\color{myteal}\selectfont{pod}}{\unskip\egroup}}}3
               {nf}{{\unskip\bgroup{\fontfamily{lmtt}\fontseries{b}\color{myteal}\selectfont{nf}}{\unskip\egroup}}}2
               {002fdc87df}{{\unskip\bgroup{\fontfamily{lmtt}\fontseries{b}\color{myteal}\selectfont{002fdc87df}}{\unskip\egroup}}}9
               {831ed4f74f}{{\unskip\bgroup{\fontfamily{lmtt}\fontseries{b}\color{myteal}\selectfont{831ed4f74f}}{\unskip\egroup}}}9
               {0f2a66}{{\unskip\bgroup{\fontfamily{lmtt}\fontseries{b}\color{myteal}\selectfont{0f2a66}}{\unskip\egroup}}}5
               {482475}{{\unskip\bgroup{\fontfamily{lmtt}\fontseries{b}\color{myteal}\selectfont{482475}}{\unskip\egroup}}}5
               {'}{{\textquotesingle}}1
               {-}{-}1
               {~}{{\centeredtilde}}1]
# subscribed Kubernetes event: kubectl get events --watch
LAST SEEN   TYPE     REASON  OBJECT
27m         Normal   Started pod/nf-002fdc87df831ed4f74f0f2a66482475
# query tags: kubectl get pod nf-002fdc87df831ed4f74f0f2a66482475 -o=jsonpath='{.metadata.labels}'
{[...] taskName":"NFCORE_RNASEQ_RNASEQ_FASTQ_FASTQC_UMITOOLS_TRIMGALORE_TRIMGALORE_WT_REP2"}
\end{lstlisting}
\caption{Example Kubernetes pod start event, API call and result (relevant fields only).\label{fig:kubernetes-trace}}
\end{subfigure}\medskip

\begin{subfigure}[t]{1\textwidth}
\captionsetup{skip=-3pt}
\begin{lstlisting}[style=customc]
[2023-12-12T16:18:11.810+0000] {scheduler_job.py:550} INFO - Sending TaskInstanceKey(dag_id='force', task_id='prepare_level2', run_id='manual__2023-12-12T16:15:47.103493+00:00', try_number=1, map_index=-1) to executor with priority 3116 and queue default
[2023-12-12T16:18:11.810+0000] {base_executor.py:95} INFO - Adding to queue: ['airflow', 'tasks', 'run', 'force', 'prepare_level2', 'manual__2023-12-12T16:15:47.103493+00:00', '--local', '--subdir', 'DAGS_FOLDER/s1/force/workflow.py']
\end{lstlisting}
\caption{Example Airflow log file entry.\label{fig:airflow-trace}}
\end{subfigure}

\caption{Exemplary excerpts from different logs and traces (colorized to show links).\label{fig:log-examples}}
\end{figure*}

% log format
The log data is written by the usermode process to a file in csv format.
The same columns are used for all log entry types. Unused columns are set to 0 or empty, e.g., the path is only used for the entry types of open and delete. Reads and writes refer to an already opened file and do not include its path again. %The columns, while still separated by commas, also have a fixed length. This enables optimizations to parse the data for analysis purposes.
\figref{fig:log-example} shows the structure of the csv file and two exemplary entries.

\subsection{Observed Limitations and Challenges}\label{sec:limitations}

We found several limitations and challenges for the eBPF monitoring to be aware of.

\subsubsection*{Unstable Signatures}
The eBPF code is attached directly to kernel functions rather than tracepoints as there are no suitable tracepoints in the VFS layer to achieve the same functionality. When signatures of functions we attach to change between kernel versions, this may break the monitoring. For example, the signature of \code{vfs_unlink} changed in kernel version 5.12. To support both kernel versions below and above that version, the correct signature can be chosen using a preprocessor directive. The more functions we attach to, the more likely a breaking change becomes. A solution using eBPF, which relies on internal kernel functions, cannot be expected to survive over many kernel versions without any maintenance.

\subsubsection*{Filter by Directory}
Another challenge is filtering by directory. The VFS layer does not receive the path of the affected file as a single string, but instead as a linked list of \code{dentry} structs, each containing a path segment starting with the filename, followed by the folder it resides in, the parent directory of that, and so forth. This makes filtering by directory names challenging in eBPF as iterating through the linked list is tricky with the restrictions the verifier puts in place for loops to hold the termination guarantee. Therefore, we currently filter by directory in the usermode process, simply omitting logs not belonging to files in directories of interest.

\subsubsection*{Amount of Data / Loss of Data} \label{sec:dataloss}
% depending on application profile and access patterns, log data can get quite large
As every single I/O access is logged, depending on the application and its I/O intensity, the log data itself can get quite large, which is an issue for detailed supercomputing monitoring solutions as well~\cite{beacon}.
% can create or increase IO bottleneck
This can cause an I/O bottleneck being created or increased by the usermode process writing the logs to storage.

% problems from bottlenecks:
Such an I/O bottleneck can degrade the performance of the application being monitored if it uses the same storage as the logs are written to.
Additionally, it can have severe consequences for the monitoring itself.
% ultimatively leads to loss of data (as eBPF cannot "wait" for usermode application)
%   1. all to file -> write speed to local disk slower than data rate of log data generation
The system may slow down the usermode part of the monitoring process. The eBPF kernel component, however, will \emph{not} slow down due to its non-blocking behavior, causing the ring buffers used for communication with the usermode part to fill up. Once full, monitoring data will get lost as the ring buffer is overwritten before it was cleared by the usermode process.
%   2. even the overhead to communicate between the eBPF kernel code and the usermode app can become bottleneck if usermode app has low priority and / or many other applications run on the node
Such a scenario can also occur even if the monitoring data is not written to a file by the usermode part but just discarded (i.e., no I/O load is caused by the monitoring). If the system is under heavy CPU load, the monitoring application may not get enough CPU time if it is not prioritized. The eBPF component, running in the respective kernel context of the applications doing the monitored I/O calls, may produce data faster than the usermode part is able to consume it. The ring buffers fill up and eventually monitoring data will be overwritten and lost.

% filters to alleviate this issue
The risk of overflowing ring buffers can be alleviated by reducing the number of events to be logged in the first place for the use case. We provide several filters implemented in the eBPF part directly so that they not only reduce the amount of data written to the log file, but also the number of entries to be transmitted through the ringbuffers.
% pid filter also for child processes
It is, for example, possible to filter by pid, user and group of the process that does the I/O call. The pid filter also supports filtering any child processes recursively. This is done by additionally attaching eBPF functions to the \code{fork} and \code{exit} tracepoints and respectively adding or removing pids accordingly in a set, which is checked by calling pid in the I/O calls.

% filesystem filter also strong
Another approach is to filter by the filesystem the I/O access happens to. By attaching to a filesystem specific open function (such as \code{ceph_open})---instead of the more general and higher-level \code{vfs_open}---only open calls in directories of that filesystem will be processed and communicated to the usermode application. Furthermore, as only actions concerning a file handle are processed, if the handle is known from a previous open call, only actions on the specified filesystems will be processed.
As the general VFS functions call these filesystem specific functions through a function pointer, they all adhere to a fixed signature, making it easy to implement the filter and use it for different filesystems.
With distributed executions of scientific workflows, the actual input and output data of the tasks is often placed in a distributed filesystem (e.g., Ceph). On many systems, the filesystem is then solely used by the scientific workflow, which makes it easy to filter them and leave out all local filesystem accesses.

% detection: BCC emits warning (on stderr) at runtime (todo: here or later?) (probably put the whole paragraph to eval)
BCC will issue warnings on \texttt{stderr} on a ring buffer overflow, which makes detecting loss of monitoring data easy.

\section{Associating Traces with Workflow Tasks}
\label{sec:associating-tasks}

For associating low-level traces, as collected by the eBPF-based monitoring described in \secref{sec:implementation}, for example, with upper-level entities such as physical and logical tasks of an SWMS, we have to bridge an information gap and find links between these levels as the entities are identified in different ways by the software components involved.

\begin{figure}[ht]
  \centering
  \includegraphics{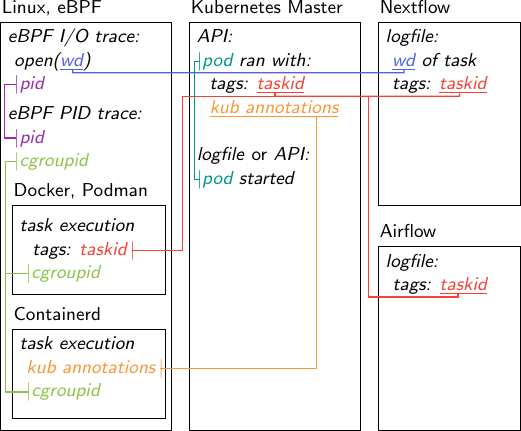}
  \caption{Architecture of mapping low-level monitoring data to upper-level identities such as tasks using information such as a task's
  individual working directory \textsf{\emph{wd}} (blue), \textsf{\emph{taskid}}
  given in tags (red), the \textsf{\emph{cgroupid}} of processes
(green), its \textsf{\emph{pid}} (violet) and Kubernetes \textsf{\emph{pod}}
(teal) and annotations \textsf{\emph{kub annotations}} (orange). Following chains of this information from component to component, the low-level trace data
  can be associated to the upper-level tasks of the workflow engine, also across machine boundaries.
  \label{fig:architecture}}
\end{figure}

\subsection{Information Gap}\label{sec:information-gap}

Depending on the particular infrastructure for workflow execution, we have two to three levels of abstraction.
On the lowest-level (leftmost in \figref{fig:architecture}), we have one or more instances of a Linux system where tasks can either be executed directly or inside a container such as Docker, Podman, or containerd. In the middle, we optionally have a resource management system, also known as an orchestrator or batch system, such as Kubernetes or Slurm. At the highest-level (rightmost in \figref{fig:architecture}), we have a SWMS such as Nextflow or Airflow.

While the lowest level identifies processes using system-local process ids (short \emph{pid}), the middle and upper level need and use more general concepts and identifiers such as container instances, physical tasks, logical tasks that the lowest level knows nothing about. When we bridge the information gap from \emph{pids} to physical tasks, the step to logical tasks and workflows can be made with means of the particular SWMS.

\paragraph*{A task can have many processes}
To associate pid-based monitoring data with tasks, we have to know which pids belong to a task. A task can be a single process, a script, or a container instance and may spawn new child processes in the operating system by itself.
% additional eBPF script that gathers cgroup id of processes and records process creations
Therefore, we keep track of the pids over time by an additional eBPF script attaching to the \code{sched_process_fork} tracepoint. It uses BCC to monitor process creations and log the parent and child \emph{pid} as well as the \emph{cgroupid} in  simple csv format (see \figref{fig:ebpf-cgroupid-trace}). Many resource management systems and containers use Linux control groups (cgroups)~\cite{cgroups} to enforce resource restrictions on processes or containers. To identify all processes that share a certain resource budget, a so called \emph{cgroupid} is assigned and associated with each process belonging to it.

\paragraph*{Tasks are not directly spawned by the SWMS}
%   1. often a container solution such as docker is used -> the "main process" of the container will not spawn as child of the swms but as child from the container daemon
Although the workflow management system is ultimately responsible for triggering task executions, the respective processes are often not created directly by it and are not simply child processes of the SWMS. With a container solution such as Docker, for example, the workflow management system sends a request to the Docker daemon, which spawns a new container process for the task. That way, the task process is a child of the Docker daemon and not from the SWMS process in the process tree.

%   2. even if process tree would be "conventionally intact", no or only weak connection to the workflows task as one task can appear multiple times (in multiple locations) in a workflow and / or one program may belong to different tasks
Even if the task processes are spawned directly by the workflow management process, i.e., the process tree would be `conventionally intact', it would not convey enough information. While deducting that a process belongs to \emph{a} task might be simple, there would still be no information available regarding to \emph{which} task exactly it belongs to. Since one program can be used in multiple tasks or one task can be used in multiple instances in the same workflow, even knowing the executed program would not suffice to make the connection.

% implicit conclusion: process tree is not enough
%So, having the process tree and \emph{cgroupids} is not enough.

\paragraph*{Tasks are distributed across several machines}

% swms often has no information about the low level pid (or properties in general) of a task as it "just" submits to docker / kubernetes
In a distributed setting, the workflow management process usually only runs on one node. Task processes are spawned using, e.g., Kubernetes, on other nodes and thus are naturally detached from the process tree of the management process.
Thus, the workflow management system has no direct knowledge of the \emph{pid} of the corresponding process.
Depending on the resource management component, the workflow management system could acquire the low level information about the started task. But for Kubernetes, for example, it is not directly possible to query the \emph{pid} or \emph{cgroupid} of a task and the investigated SWMSs do not provide such information.

% ultimatively only source for the knowledge about the tasks is the workflow management system
On the other side, logical and physical tasks are only known by the workflow management system itself. It starts or queues the tasks and possesses the information of how they fit in the workflow.
% swms may pass information on to other components (e.g. kubernetes), but no uniform way to query such information (e.g. kubernetes annotations of nextflow differ from those of airflow)
The corresponding information may be passed on to the resource management that schedules the execution. That happens, for example, when Nextflow or Airflow queue tasks with Kubernetes. They annotate pods that execute workflow tasks with
specific tags (see \figref{fig:architecture}) that make them identifiable. The tags also state which concrete task the pod executes and to which workflow it belongs---both provided by the workflow management system as unique ids (see \code{name} in \figref{fig:nextflow-trace} for Nextflow and \code{dag_id} and \code{task_id} in \figref{fig:airflow-trace} for Airflow).

\paragraph*{Pids seen inside tasks may be misleading}
% wrapper script (as done e.g. by nextflow to collect summarized task statistics) not possible as querying a pid from within the container usually does not provide the actual node-local pid, but just a container-local pid
When a task is executed, in a Kubernetes pod, locally or through other means of execution, it may be wrapped by a bash or job script of the workflow management system to gather summarized execution statistics of the task. While it would be trivial to expand such a wrapper script to also gather the \emph{pid}, that information will often be useless. If the execution happens in a containerized environment, the gathered \emph{pid} often is local to the container the task runs in. It cannot be easily matched to the actual \emph{pid} that monitoring tools outside of the task's container perceive.

\subsection{Bridging the Gap Using Special Nextflow Behavior}\label{sec:nextflow-wd}

% each task gets a unique folder as working directory inside the workflows working directory - the unique working directory appears in the nextflow logs next to the tasks name
In Nextflow, every task gets its own working sub-directory within the workflow's folder. That working directory is logged with its task to the workflow's execution log when the task finishes (see \figref{fig:nextflow-trace}).
% however, tasks can also access the files of other tasks working directories (mostly reading them as their input, but writing is not prevented technically)
Although having separate directories, tasks may access files in directories of other tasks when they read the output of a preceding task. Actually, tasks could even write to files in other task's folders---Nextflow does not prevent that.
% nextflow has some special files per task in its directory that are exclusively accessed by that task - file accesses are logged anyway together with the pid -> acquiring the "main pid" of a task
Apart from task-specific files, Nextflow puts additional files into the task's directory such as a wrapping bash script to run the task or a file with the return code of the task when it finishes. These files are not accessed by other tasks.
% since tasks run in containers and every container has its own cgroup, all pids in a cgroup belong to one task -> use the cgroup and process creation log to get all pids of a task
When Nextflow runs tasks in Docker containers, either directly (for single node executions) or via Kubernetes (for distributed executions), all processes that execute within a container---and therefore belong to one task of the workflow---share the same \emph{cgroupid}.
% task -(nf log)-> working directory -(access log, special file access)-> task main pid -(cgroup and process creation log)-> all task pids
When a process accesses one of the Nextflow-specific files in a task-specific directory, it can then be concluded that the corresponding \emph{cgroupid} belongs to that task. All file accesses of all processes that share this \emph{cgroupid} are then categorized to belong to that task.

Without containers, the association can be made by inspecting the process tree after the process accessing the unique job script file has been found. Either a shared unique \emph{cgroupid} can be found, or the all pids that belong to that task can be derived to associate the corresponding monitoring data with the task.

\subsection{Bridging the Gap Using Docker and Kubernetes}

In case of Kubernetes, a script can subscribe to events to get notified whenever a new pod is started (see \figref{fig:kubernetes-trace}).
The entry does neither include the main \emph{pid} of the pod, its \emph{cgroupid} nor workflow tag information.
% if a relevant (meaning belonging to a workflow) pod is started, grab its pid / cgroupid
To make this association, newly spawned pods can then be checked on their workflow system's tag through the Kubernetes API (see \figref{fig:kubernetes-trace}) and/or their tags and \emph{cgroupid} can be queried on the executing node by a self-written script using the API of Docker, Podman or containerd.
As outline in the previous section, knowing the \emph{cgroupid} leads to all \emph{pids} related to a task and the file accesses can be associated to this task.
% another approach would be to regularly query all running pods, check which are annotated and check them as described
Instead of watching the Kubernetes logs, all pods can be queried---regularly or triggered by spawns reported by our eBPF pid tracer (see \secref{sec:information-gap})---on their nodes to check for workflow tags and \emph{cgroupid}s.
%   that can lead to misses as short running tasks may both start and end between two queries
All these approaches bear the risk of `missing' short-lived tasks that are already finished before the corresponding query happens.

\subsection{Bringing Distributed Log Data Together}
For analysis, all traces and logs have to be brought together. We foresee a directory per involved host, one for the SWMS' log, and one for the resource management logs. Either the traces and logs are written to these locations on a shared file system or they are copied together after the workflow run.

\section{Demonstration}
\label{sec:evaluation}

% workflows: nf-core workflows in "test" profile (uses test data from the repository)
% swms: nextflow, association using the tasks working directories as described above
To demonstrate the practical applicability of our approach, we implemented the described eBPF approach for monitoring, the Nextflow specific association technique and used workflows from the nf-core repository~\cite{nfcore} for the analysis. We executed them with their test profile in which they provide their own data. That proved sufficient to demonstrate the ability of our monitoring approaches to obtained detailed I/O monitoring data over time per workflow task.

\subsection{Setup}

Two setups were used for the evaluation. Nextflow was used as workflow management system if not stated otherwise.

\paragraph*{(1) Local setup}
Workflow execution with Nextflow 23.10.1 was done locally inside a VMware virtual machine (VM) using Docker without a separate resource manager such as Kubernetes or Slurm. While the host ran Windows 10, the VM used Debian 11 as operating system, Linux kernel 5.10.0. The local disk, which was used for the workflow's working directory, used an ext4 filesystem. It was a vmdisk image that was saved on an SSD using NTFS mounted to the host system. Most workflow executions were done on this system.

Processes and tasks were associated via the tasks' unique working directories as described in \secref{sec:nextflow-wd} (see also \figref{fig:architecture}).

\paragraph*{(2) Cluster setup}
% tu cluster: monitoring per node, ceph
A Kubernetes cluster of four nodes, each running Ubuntu 20.04 LTS with Docker as container solution was used as an alternative. Here, the workflows, running through Nextflow 23.04.2, used a Ceph filesystem (based on the corresponding kernel module, not the FUSE-based client) to exchange data. The monitoring process as well as the \emph{pid} and \emph{cgroupid} tracer were deployed on every node using a Kubernetes daemonset with their data being written to a node-local storage.
% on clusters monitoring was limited to ceph (as described above), locally no limitation
As the nodes were shared and other applications were running on them too, the local storage was initially overloaded by the amount of monitoring data, causing log entry loss as described in \secref{sec:dataloss}. 
By limiting the I/O monitoring explicitly to the Ceph filesystem (as described in \secref{sec:limitations}) this bottleneck was avoided.

\paragraph*{Data analysis}
% script processed each monitoring log together with the .nextflow.log file for association one after the other to gather the data for the graphs and statistics
A script analyzes the monitoring data and associates them with the Nextflow tasks. It receives the \texttt{.nextflow.log} file of a workflow execution and a list of directories containing the I/O monitoring data and process creation CSV files. Each directory contains the logs from one node, so that \emph{pids} and \emph{cgroupids}, which are not unique across multiple nodes, can each be handled separately per node. The script can therefore handle the execution traces of any number of nodes including single-node local executions.

\subsection{Results}

\begin{figure}[t!]
  \includegraphics[width=\linewidth]{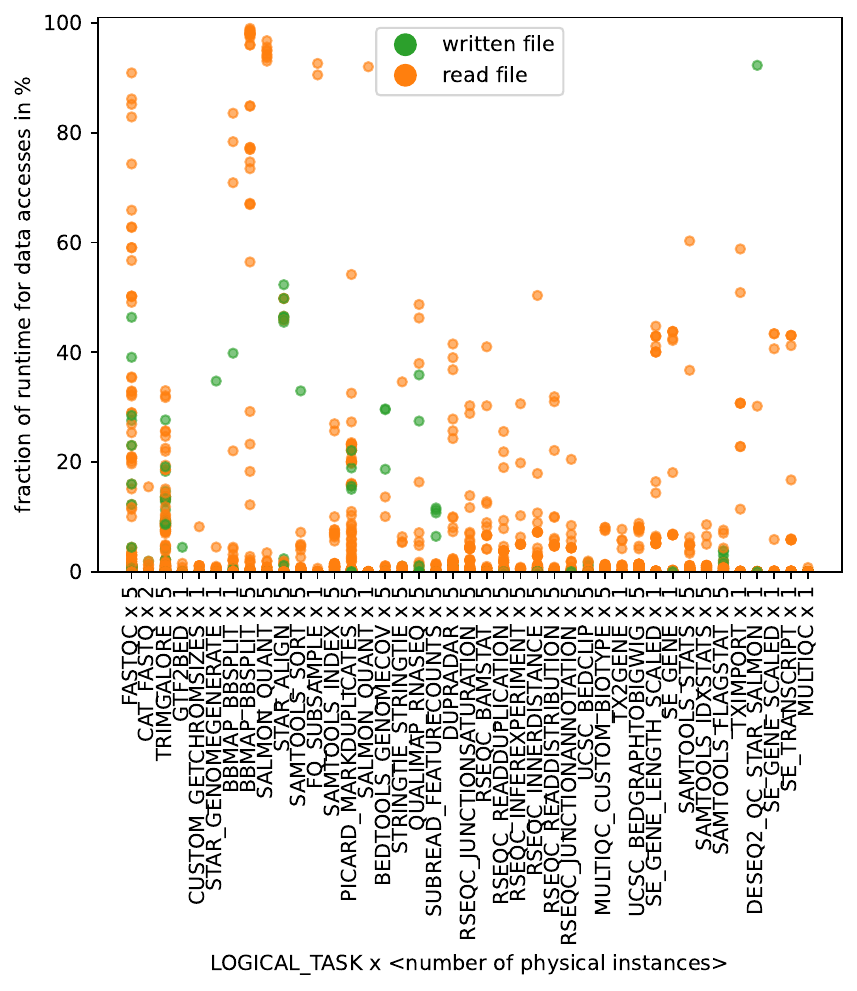}
  \caption{Duration between first and last access of a file from a
task relative to the task's runtime. Analyzed from a execution of the
\emph{rnaseq} workflow of \emph{nf-core} in the \emph{test} profile. A larger fraction
for a file likely indicates a streaming-like access while a small
fraction represents more bulky file access of a task.
  \label{fig:bulkishness}}
\end{figure}

The monitoring data obtained with our approach enables more detailed and elaborate analysis than the summary data that Nextflow provides in its trace reports.
Besides all files, file sizes, and which task accessed them, the concrete access patterns and their distribution over the task's runtime can be investigated, for example. \figref{fig:bulkishness} shows how much time passes between the first and last access of a file that is read or written by a task relative to this task's total runtime for all tasks of the \emph{rnaseq} workflow. A value close to 100\,\% indicates that the file is read or written over the whole runtime while a low value shows a more `bulky' access that happens in a relatively short time. Overall, most accesses are finished fast here due to many small configuration and log files and the relatively small data set contained in the corresponding test profile.

\begin{figure}[t!b]
  \includegraphics[width=\linewidth]{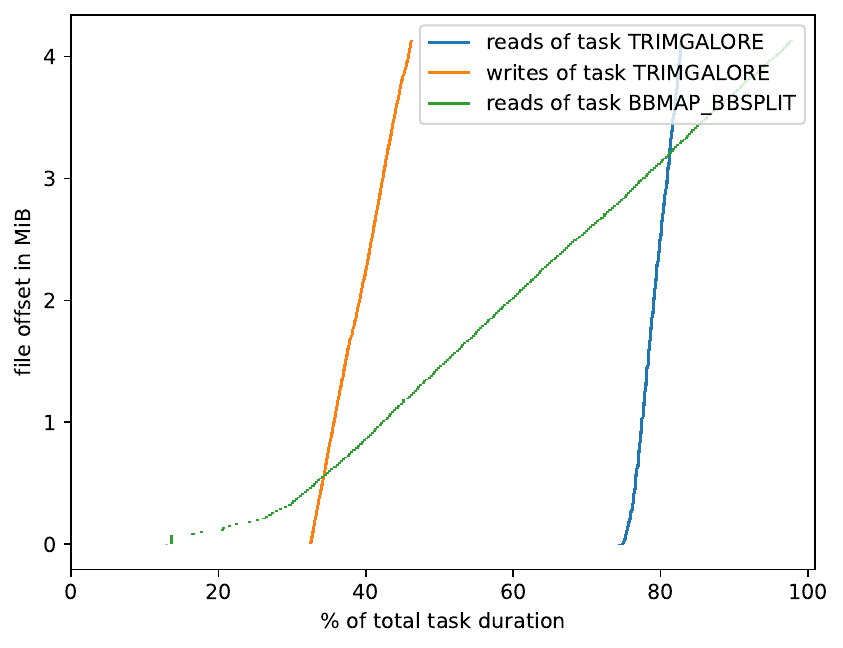}
  \caption{File \code{WT_REP1_2_val_2.fq.gz} is accesses by two tasks. \emph{TRIMGALORE} creates and writes it between 32 and 42\% of its runtime of 9.91s in total. The same task also reads the file back once later. A succeeding task, \emph{BBMAP\_BBSPLIT} reads it over almost its whole lifespan of 24.16s.
  \label{fig:accesses}}
\end{figure}

\figref{fig:accesses} shows the accesses, both reads and writes, to a specific file in more detail. The plot consists of vertical bars, one per read or write. The lower end of each bar starts at the operation's file offset and its height represents the number of bytes accessed in the request.

The runtime overhead of our eBPF-based monitoring is negligible and below the normal variance of the overall workflow and task execution times.

\section{Related Work}

% Access pattern claim is a bit flimsy, there's no work that directly explores that so far
In general, the topic of data provenance and the discovery of access patterns of scientific workflows, or systems that do end-to-end monitoring of low-level information in the context of scientific applications resemble this work or have similarities.

%% moved here for better figure placement

% Here the difference is that they collect performance data using existing specialized tools and that they are both HPC oriented, while we are commodity cluster focused
To correlate, store and make fine-grained provenance information available, particularly for the Pegasus SWMS, performance data can be gathered using specialized tools that are well-established in the HPC community~\cite{papadimitriou_end--end_2021}. Similarly, with an adapter-based framework that processes and integrates events from various systems, these can be made available in a DBMS~\cite{souza_towards_2023} while processing multiple workflows in parallel. Both systems contain a clear system structure that integrates the information and stores it in dedicated storage, however, both works focus on systematic correlation of the data to scientific workflows, while the low-level information is left to be handled by external tools.

% Data Provenance in HPC
% PROV-IO+: A Cross-Platform Provenance Framework for Scientific Data on HPC Systems
The importance and difficulty of integration of monitoring data is highlighted also in \cite{han_prov-io_2023} with a focus on HPC. By working closely with domain scientists, they identify their needs and implement a framework that can track either data lineage using an explicit linking of data objects by mirroring the API of the underlying HDFS storage layer, or ad-hoc performance metrics by implementing dedicated system calls. The framework and the model presented are versatile but still require manual work for a particular use-case, and while cross-platform they focus on requirements in the context of HPC.

% Data provenance in Cloud
% Progger: An Efficient, Tamper-Evident Kernel-Space Logger for Cloud Data Provenance Tracking
Monitoring of scientific workflows needs similar efforts in a cloud environment~\cite{ko_progger_2014}. Here, kernel-space logging allows automatic linking of data accesses to processes. However, the focus is more on the security aspect and not on the requirements of scientific workflow development and usage.

To address the need for low-level information retrieval and enabling the discovery of inefficient resource usage of JVM based applications, a filesystem framework tracing detailed I/O usage and attribution in big data software stack scenarios
was developed~\cite{iotracing} resembling the monitoring approach of a FUSE overlay filesytem. The association to the JVM-based applications is done by wrapping relevant I/O methods of core Java classes and adding logging on that layer as well.

In all works, the desire for automated coupling of monitoring information with higher-level abstractions of each software and infrastructure stack is an important challenge. Additionally, most works aim to particularly track I/O monitoring data, indicating its importance, regardless of the application.

\section{Conclusion}
% implementation conclusion
% detailed low level monitoring of workflows not as easy as for a single application
Low-level monitoring solutions for scientific workflows face the lack of a unified method to link low-level monitoring data with abstract workflow tasks.
% instead: reliance on "intermediate" components (kubernetes) or specifics of the workflow management system
Instead, we rely on either `intermediate' components between the workflow management system and the actual execution (such as Kubernetes and Docker) or on specifics of the SWMS such as unique working directories per task and specific files as described for Nextflow.

We established links between different information sources to associate the low-level monitoring data with upper-level entities. That it was possible at all in our scenarios, was probably more luck than intended by the design of the used software systems. For example, in the case of Airflow, we do not see a mean to perform this association when the tasks are not executed via an orchestrator such as Kubernetes but container instances would be issued directly. Logs of different system components can play a vital role for unforeseen use cases and application analysis. In the best case, they contain both upper-level and low-level identifiers in their entries that allow to bridge the information gap.

% link to code
The code of our monitoring tool, a Python script realizing the association and the task-based analysis of monitoring data are at \url{https://github.com/CRC-FONDA/workflow-monitoring}.

\section*{Acknowledgements}

We thank Sören Becker from TU Berlin who helped us to implement our monitoring solution as a container and provided the Kubernetes daemonset for monitoring in a distributed setting. We thank TU Berlin, HU Berlin, TU Darmstadt and ZIB's IT department for providing compute resources.

\bibliographystyle{IEEEtranS}
\bibliography{main}

\end{document}